\documentclass[aip,pop,reprint]{revtex4-1}
\pdfoutput=1
\usepackage[utf8]{inputenc}
\usepackage[T1]{fontenc}
\usepackage{mathtools}
\usepackage{amsfonts}
\usepackage{amssymb}
\usepackage{graphicx}
\usepackage{siunitx}
\usepackage{lmodern}
\usepackage{physics}
\usepackage{microtype}
\usepackage{natbib}

\usepackage{hyperref}
\usepackage{epstopdf}
\hypersetup{
    linkcolor=blue,
    citecolor=blue,
    urlcolor=blue,
    colorlinks=true,
    pdfauthor={},
    pdftitle={}
}
\newcommand{\ion}{\textup{i}}

\newcommand{\smax}{\textup{max}}

\newcommand{\plasm}{\textup{p}}

\newcommand{\yb}{y_\mathrm{b}}
\newcommand{\rb}{r_\mathrm{b}}
\newcommand{\bunch}{\textup{B}}
\newcommand{\laser}{\textup{L}}

\newcommand{\asquare}{{\left\langle \vb{a}^2 \right\rangle}}

\begin{document}
\title{Bubble regime of plasma wakefield in 2D and 3D geometries}
\author{A.\,A. Golovanov}
\author{I.\,Yu. Kostyukov}
\affiliation{Institute of Applied Physics RAS, 603950 Nizhny Novgorod, Russia}

\begin{abstract}
    Considering the popularity of two-dimensional particle-in-cell simulations, a 2D model of plasma wakefield in the strongly nonlinear (bubble) regime in transversely non-uniform plasma is developed.
    A differential equation for the boundary of the bubble in the 2D geometry is obtained, its analytic solution is derived.
    2D particle-in-cell simulations are used to confirm the validity of our model.
    The results are compared to the bubble in the realistic 3D geometry.
    For uniform plasma, it is shown that the 2D bubble is elongated and has stronger focusing forces, while the structure of the accelerating field remains completely unchanged.
    A method of generating a quasi-2D bubble in the realistic three-dimensional geometry is proposed.
\end{abstract}
\maketitle

\section{Introduction}

Nowadays, a lot of attention is drawn to plasma acceleration methods. \cite{Esarey2009RevModPhys, Kostyukov2015UFN}
Compared to conventional radio-frequency linacs, plasma accelerators can provide orders of magnitude higher acceleration gradients.
The main idea of these methods is to use a driver to excite a plasma wake wave whose longitudinal electric field can be used to efficiently accelerate co-propagating charged particles.
A short intense laser pulse \cite{Tajima1979laser} or a relativistic electron bunch \cite{Rosenzweig1988experimental} can be used as a driver, corresponding to laser-wakefield acceleration (LWFA) and plasma-wakefield acceleration (PWFA), respectively.
The experiments on plasma acceleration demonstrate acceleration gradients of tens of gigavolts per meter.
For example, in the leading LWFA experiments, accelerated electrons with the energy of \SI{4.2}{\GeV} for the acceleration distance of \SI{9}{cm} have been obtained.\cite{Leemans2014PRL}
For PWFA, the energy increase from \SI{42}{\GeV} to more than \SI{80}{\GeV} over the distance of \SI{85}{cm} has been observed.\cite{Blumenfeld2007Nature}

For sufficiently intense laser pulses or sufficiently dense electron bunches, the driver interacts with plasma in the strongly nonlinear regime, leading to the formation of a near-spherical plasma cavity (a bubble) free of plasma electrons. \cite{Pukhov2002Bubble}
On the boundary of this bubble, a thin electron sheath shielding the cavity from the surrounding plasma is formed.
In this regime, self-injection is possible,\cite{Froula_2009_PRL_103_215006} i.\,e. electrons from the background plasma are trapped and accelerated in the bubble, which is commonly used in experiments.

There have been significant advancements in the theoretical description of the bubble regime over the recent years.
A simple model in which the bubble is assumed ideally spherical can be used to qualitatively describe the bubble regime. \cite{Kostyukov_2004_PoP_11_115256}
A more detailed phenomenological model makes it possible to describe the boundary of the bubble with a differential equation. \cite{Lu_2006_PoP_13_056709}
This phenomenological model has also been generalized for plasmas with non-uniform transverse profiles \cite{Thomas_2016_PoP_23_053108, Golovanov_2016_QE_46_295}, and it is also capable of describing beam loading effects \cite{Tzoufras_2009_PoP_16_056705, Golovanov_2016_PoP_23_093114} (i.\,e. the influence of accelerated electron bunches on the bubble).
In the scope of the model, explicit expressions for the electromagnetic field components both inside and outside the bubble can be obtained. \cite{Golovanov_2017_PoP_24_103104}
Furthermore, scaling laws based on the similarity theory have been obtained for the bubble regime both in uniform plasmas \cite{Gordienko_2005_PoP_12_043109} and plasmas with channels. \cite{Pukhov2014Channel}

Despite the achievements in the theoretical description, the phenomenological and not self-consistent nature of current models limits their use for the description of LWFA and PWFA.
Numerical simulations with the particle-in-cell (PIC) method remain the most general way of studying laser--plasma and beam--plasma interactions. \cite{Pukhov2016PIC}
Being based on fundamental equations, such simulations can self-consistently capture most of the relevant physical effects and can be used as a tool for ``numerical experiments.''
However, due to their nature, full 3D PIC simulations often require immense computational resources, which can be prohibitive for many problems.
For simulations of laser--plasma interaction, distributed machines with hundreds of gigabytes of RAM are often necessary.
It is also not unusual for full LWFA and PWFA simulations to take weeks of time on modern multi-processor systems.
This can significantly limit the possibility of performing series of simulations for a wide range of parameters, and that is why simpler simulation methods are often used.
One of them is 2D PIC simulations in which a two-dimensional grid is used instead of a realistic 3D grid, which significantly reduces the amount of required resources.
From the physics point of view, it corresponds to a driver and wakefield infinitely long and completely uniform in one direction.
Despite the fact that this geometry is different from the realistic one, such simulations are actively used in theoretical studies, e.\,g. in Refs. \onlinecite{Zhang_2015_PRL_114_184801, Sahai_2017_arxiv_1704.02913, Shaw_2014_PPCF_56_084006, Petrillo_2008_PRSTAaB_11_070703, Papp_2018_arxiv_1801.04093, Williamson_2017_arxiv_1712.00255}.
Because of that, understanding the difference in the structure of the wakefield between 2D and 3D geometries is important.

In this paper, we develop a model of strongly nonlinear wakefield in the 2D Cartesian geometry.
The developed model is similar to the model of the bubble in the 3D geometry. \cite{Lu_2006_PoP_13_056709, Thomas_2016_PoP_23_053108, Golovanov_2016_QE_46_295}
The paper is structured as follows.
In Sec.~\ref{sec:wakefieldGeneral}, we provide basic equations for the description of the wakefield.
Then, in Sec.~\ref{sec:motion}, we describe the trajectories of electrons in the wakefield.
The model of the bubble in the 2D geometry is introduced in Sec.~\ref{sec:bubbleModel}.
Based on this model, an equation for the bubble boundary is obtained and solved analytically in Sec.~\ref{sec:bubbleEquation}.
The theoretical results are compared to the results of 2D PIC simulations.
Finally, in Sec.~\ref{sec:quasi2d}, the possibility of creating a bubble similar to the 2D bubble in the realistic 3D geometry is considered.

\section{Equations for the wakefield}
\label{sec:wakefieldGeneral}

Let us consider a driver (an electron bunch or a laser pulse) propagating in fully ionized plasma along the $x$ axis and exciting wakefield in the strongly nonlinear regime.
We assume the 2D geometry in which the driver is infinite in the $z$ direction, and therefore all values are independent of $z$.
The plasma density $n(y)$ depends only on the transverse coordinate $y$.
This allows us to consider plasmas with different types of channels in addition to uniform plasma.
Both the driver and the plasma density distributions are assumed to be symmetric about the $y=0$ plane.
In this paper, we use unitless values in which charges are normalized to $e$, masses to $m$, time to $\omega_\plasm^{-1}$, coordinates to $c/\omega_\plasm$, densities to $n_\plasm$, electric and magnetic fields to $mc\omega_\plasm/e$.
Here, $e>0$ is the elementary charge, $m$ is the electron mass, $n_\plasm$ is the typical electron number density (for example, for plasma channels, it could be the density far outside the channel), $\omega_\plasm = (4\pi e^2 n_\plasm/m)^{1/2}$ is the corresponding typical plasma frequency.

It is convenient to describe the electromagnetic field with the scalar potential $\varphi$ and the vector potential $\vb{A}$.
If we take into account the 2D geometry and the symmetry with respect to $y=0$, only three non-zero components of the electromagnetic field exist
\begin{align}
    &E_x = -\pdv{A_x}{t} - \pdv{\varphi}{x}, \quad E_y = -\pdv{A_y}{t} - \pdv{\varphi}{y},\\
    &B_z = \pdv{A_y}{x} - \pdv{A_x}{y}.
\end{align}
Both the fields and the potentials depend on time $t$ and coordinates $x$ and $y$.
However, it is typical that the structure of the wakefield changes slowly during its propagation through plasma, so the dependence on $t$ and $x$ can be replaced with the dependence on $\xi = t - x$, which is called ``the quasistatic approximation''.
In this case, the phase velocity of the wakefield is assumed to be equal to the speed of light (1 in unitless values).
Under this approximation, all derivatives with respect to $x$ and $t$ are replaced with derivatives with respect to $\xi$
\begin{align}
    &E_x = \pdv{\Psi}{\xi}, \quad E_y = -\pdv{\Psi}{y} + B_z,\\
    &B_z = -\pdv{A_y}{\xi} - \pdv{A_x}{y}.
\end{align}
Here, we have introduced the wakefield potential $\Psi = \varphi - A_x$.
For the potentials, we use the Lorenz gauge
\begin{equation}
    \pdv{A_y}{y} = - \pdv{\Psi}{\xi}.
\end{equation}
Under the symmetry constraints, it leads to 
\begin{equation}
    A_y = - \int_0^y {\pdv{\Psi}{\xi} \dd{y'}},
\end{equation}
thus leaving only $\Psi(\xi, y)$ and $A_x(\xi, y)$ as independent potentials.
The Maxwell's equations for these potentials in coordinates $(\xi, y)$ reduce to 
\begin{equation}
    \pdv[2]{\Psi}{y} = J_x - \rho,\quad \pdv[2]{A_x}{y} = - J_x.
\end{equation}
Their solutions are
\begin{align}
    &\Psi = -\int_y^\infty \dd{y'} \int_0^{y'} {(J_x - \rho)\dd{y''}},\label{eq:PsiGeneral}\\
    &B_z = \int_0^y {\left(\pdv[2]{\Psi}{\xi} + J_x\right) \dd{y'}}.\label{eq:BzGeneral}
\end{align}
These equations allow us to calculate the distributions of $\Psi$ and $B_z$ if we know the distributions of sources $J_x$ and $J_x - \rho$.

Knowing the wakefield potential $\Psi$ is extremely important for studying the acceleration of particles in the wakefield.
If we consider a relativistic particle moving predominantly along the $x$-axis ($\abs{p_y} \ll p_x$), then the forces acting on such a particle are
\begin{align}
    &F_x \approx - E_x = - \pdv{\Psi}{\xi}, \label{eq:FxGeneral} \\
    &F_y \approx - E_y + B_z = \pdv{\Psi}{y}. \label{eq:FyGeneral}
\end{align}
These forces depend only on the wakefield potential, therefore its distribution fully determines the motion of accelerated relativistic particles.
In order to calculate this distribution, dynamics of plasma have to be considered.

\section{Motion of plasma electrons}
\label{sec:motion}

The most general description of collisionless plasmas in the electromagnetic field is given by the kinetic Vlasov equations for plasma components in which the electromagnetic field is treated self-consistently and depends on the plasma distribution. \cite{vlasov1968vibrational} 
According to the method of characteristics, this kinetic approach is equivalent to the solution of motion equations for test particles in the self-consistent fields.
In the $(\xi, y)$ coordinates, the equations of motion for electrons are
\begin{align}
    &\dv{p_x}{t} = - E_x - \frac{p_y B_z}{\gamma} - \frac{1}{2\gamma} \pdv{}{x} \asquare,\label{eq:pxEquation}\\
    &\dv{p_y}{t} = - E_y + \frac{p_x B_z}{\gamma} - \frac{1}{2\gamma} \pdv{}{y} \asquare,\\
    &\dv{\xi}{t} = 1 - \frac{p_x}{\gamma}, \quad \dv{y}{t} = \frac{p_y}{\gamma}. \label{eq:coordinateEquations}
\end{align}
where $\vb{a} = e\vb{E}_\laser/(mc\omega_\laser)$ is the dimensionless amplitude of the laser electric field, $\omega_\laser$ is the laser frequency,
\begin{equation}
    \gamma = \sqrt{1 + \vb{p}^2 + \asquare}
    \label{eq:gammaDef}
\end{equation}
is the Lorentz factor of an electron.
Here, we use the ponderomotive description of the laser pulse. \cite{Mora_1997_PoP_4_010217}
In this case, the field of the laser pulse is not taken into account in the Maxwell's equations and vectors $\vb{E}$ and $\vb{B}$, and the influence of the laser pulse on plasma electrons is determined by the ponderomotive force.

System of equations \eqref{eq:pxEquation}--\eqref{eq:coordinateEquations} can be described by a Hamiltonian
\begin{equation}
    H(-\xi, y, P_x, P_y) = \gamma - P_x - \varphi,
\end{equation}
where $\vb{P} = \vb{p} - \vb{A}$ are canonical momenta.
As $\varphi$ and $\vb{A}$ do not depend explicitly on time in the $(\xi, y)$ coordinates, the value of the Hamiltonian is conserved on trajectories.
For electrons initially at rest (thermal motion is neglected), this value is $H = 1$.
Hence, on the electron trajectories,
\begin{equation}
    \gamma - P_x - \varphi = \gamma - p_x - \Psi = 1.
    \label{eq:constantOfMotion}
\end{equation}
Therefore,
\begin{equation}
    \dv{\xi}{t} = \frac{\gamma - p_x}{\gamma} = \frac{1 + \Psi}{\gamma}.
\end{equation}
As $\dv*{\xi}{t}$ is always positive, $\xi(t)$ is a monotonous function.
Therefore, $\xi$ can be used instead of $t$ as a parameter for the electron trajectories.
Then, the equations for the transverse motion become
\begin{align}
    &\dv{y}{\xi} = \frac{p_y}{1+\Psi},\\
    &\frac{1 + \Psi}{\gamma}\dv{p_y}{\xi} = - \pdv{\Psi}{y} - \frac{1 + \Psi}{\gamma} B_z - \frac{1}{2\gamma} \pdv{}{y} \asquare.
\end{align}
Using Eqs.~\eqref{eq:gammaDef} and \eqref{eq:constantOfMotion}, we can find $\gamma$ through the other values as well,
\begin{equation}
    \gamma = \frac{1 + (1+\Psi)^2 + p_y^2 + \asquare}{2(1+\Psi)}.
\end{equation}

Finally, the following second-order equation for an electron trajectory $y(\xi)$ can be obtained
\begin{multline}
    \dv{\xi} \left[ (1 + \Psi) \dv{y}{\xi} \right] = - \frac{1}{2(1+\Psi)} \pdv{y} \asquare + \\ 
    + \left[\frac{1 + (1+\Psi)^2}{2(1+\Psi)^2} + \frac{1}{2} \qty(\dv{y}{\xi})^2 \right] \pdv{\Psi}{y} - B_z.
    \label{eq:electronTrajectory}
\end{multline}

A similar equation can be obtained for the ion trajectories.
However, as ions are much heavier than electrons, their motion in the bubble regime can usually be neglected.
Because of this, we consider them immobile.
Hence, their charge density $\rho_\ion(y)$ is determined only by the plasma profile $n(y)$, and their current density $\vb{J}_\ion = 0$.

In principle, self-consistent solution of Eqs.~\eqref{eq:PsiGeneral}, \eqref{eq:BzGeneral}, \eqref{eq:electronTrajectory} is required in order to properly describe the excited wakefield.
However, a simpler phenomenological model can be used in the case of strongly nonlinear wakefield.
This model is described in the next section.

\section{Model of the bubble}
\label{sec:bubbleModel}

Based on the properties of the bubble regime observed in particle-in-cell simulations, the model of the bubble in the two-dimensional case can be chosen similar to the 3D model by \citet{Golovanov_2017_PoP_24_103104}
We assume that there are no plasma electrons inside the bubble, while on its boundary determined by a function $\yb(\xi)$ there is a thin electron sheath of constant width $\Delta$.
Under this assumption, the source $J_x - \rho$ for the bubble modeled as
\begin{equation}
    J_x - \rho = \begin{dcases}
    -\rho_\ion(y),& \abs{y} < \yb(\xi),\\
    S_0(\xi) g\left(\frac{\abs{y} - \yb(\xi)}{\Delta}\right), &\abs{y} > \yb(\xi).
    \end{dcases}
\end{equation}
In this model, the space is split into two regions by curves $\pm \yb(\xi)$ corresponding to the boundary of the bubble.
Inside the bubble, only plasma ions contribute to $J_x - \rho$, as there are no plasma electrons inside.
Relativistic electron bunches (either a driver or a witness) do not contribute to $J_x - \rho$ either, because their velocity $v_x \approx 1$, and thus
\begin{equation}
    J_{x,\bunch} - \rho_\bunch = (v_x-1) \rho_\bunch \approx 0.
\end{equation}
An arbitrary function $g(X)$ describes the shape of the electron sheath on the boundary of the bubble.
Far outside the bubble, for $\abs{y} \gg \yb$, plasma should remain unperturbed, therefore $g(X)$ must tend to zero.
For example, exponential $g(X) = \exp(-X)$ and rectangular $g(X) = \theta(1-X)$ profiles have been used in previous 3D models. \cite{Lu_2006_PoP_13_056709, Tzoufras_2009_PoP_16_056705}
By multiplying $\Delta$ and $S_0(\xi)$ by constants, we can always normalize this function in a way that
its moments $M_0(0) = M_1(0) = 1$, where the moments are defined as
\begin{equation}
    M_n(X) = \int_X^\infty {g(X')\dd{X'}}.
\end{equation}
To simplify the calculations, we assume that $g(X)$ is normalized.

In order for the indefinite integral in Eq.~\eqref{eq:PsiGeneral} to converge, $\int_0^\infty (J_x - \rho) \dd{y} = 0$ is required, which allows us to find
\begin{equation}
    S_0(\xi) = \frac{S_\ion(\yb(\xi))}{\Delta},
\end{equation}
where the function
\begin{equation}
    S_\ion(y) = \int_0^y {\rho_\ion(y') \dd{y'}}
\end{equation}
is determined by the plasma profile.
Therefore, the function $\yb(\xi)$ fully determines the source $J_x - \rho$ if the properties of plasma and the electron sheath are postulated.

Knowing $J_x - \rho$, we can calculate $\Psi$ using Eq.~\eqref{eq:PsiGeneral}.
For $\abs{y} < \yb$, the resulting wakefield potential is
\begin{equation}
    \Psi(\xi, y) = \int_y^{\yb} S_\ion(y') \dd{y'} + \Delta S_\ion(\yb).
\end{equation}
According to Eqs.~\eqref{eq:FxGeneral}, \eqref{eq:FyGeneral}, the forces acting on relativistic particles in this potential are
\begin{align}
    &F_x(\xi) = -\left(S_\ion(\yb) + \Delta \rho_\ion(\yb) \right)\dv{\yb}{\xi},\\
    &F_y(y) = -S_\ion(y).
\end{align}
Similarly to the 3D axisymmetric case,\cite{Lu_2006_PoP_13_056709} the longitudinal force depends only on the longitudinal coordinate, while the transverse force depends only on the transverse coordinate.
As expected, the transverse force is always focusing for electrons.
However, the amplitude of this force is different in the 2D case.
For example, if we consider uniform plasma ($S_\ion(y) = y$), the focusing force in the 2D geometry $F_y = -y$ remains linear but is two times larger than the force in the 3D geometry $F_r = - r / 2$.
This means that electrons in 2D simulations will experience a stronger focusing force than in corresponding 3D simulations.
As this force is responsible for the transverse betatron oscillations and resulting betatron radiation of electrons,\cite{Kostyukov_2003_PoP_10_124818} this change may significantly influence the spectrum of betatron radiation observed in simulations.

In order to find the longitudinal field $E_x(\xi)$ and the corresponding longitudinal force $F_x$, we need to know the shape of the bubble's boundary $\yb(\xi)$.
As it is known from the previous 3D models,\cite{Lu_2006_PoP_13_056709, Thomas_2016_PoP_23_053108} this shape can be self-consistently found.
The corresponding calculations for the 2D case are described next.

\section{Equation for the bubble's boundary}
\label{sec:bubbleEquation}

As electrons move in the electron sheath around the bubble, the boundary of the bubble $\yb(\xi)$ at the same time serves as the innermost electron trajectory.
Therefore, Eq.~\eqref{eq:electronTrajectory} for an arbitrary electron trajectory is valid for the boundary $\yb(\xi)$ as well.
In order to use this equation, the values of the wakefield potential and its derivatives at $y = \yb$ are required.
They are
\begin{align}
    &\Psi(\xi, \yb(\xi)) = \Delta S_\ion(\yb(\xi)),\\
    &\pdv{\Psi}{y} (\xi, \yb(\xi)) = - S_\ion(\yb(\xi)),\\
    &\pdv{\Psi}{\xi} (\xi, \yb(\xi)) = \left(S_\ion(\yb) + \Delta \rho_\ion(\yb)\right) \dv{\yb}{\xi}.
\end{align}
Also, the magnetic field $B_z(\xi, \yb)$ is needed; it can be calculated from Eq.~\eqref{eq:BzGeneral}
\begin{multline}
    B_z(\xi, \yb) = \int_0^{\yb} {J_x(\xi, y')\dd{y'}} + \\
    + \yb \left[(\rho_\ion + \Delta \rho'_\ion) \qty(\dv{\yb}{\xi})^2 + (S_\ion + \Delta \rho_\ion) \dv[2]{\yb}{\xi}\right].
\end{multline}

If we substitute all of these functions into Eq.~\eqref{eq:electronTrajectory}, we obtain the equation describing the boundary of the bubble
\begin{equation}
    A(\yb) \dv[2]{\yb}{\xi} + B(\yb) \qty(\dv{\yb}{\xi})^2 + C(\yb) = \lambda + L. \label{eq:bubbleEquationGeneral}
\end{equation}
This second-order ordinary differential equation shows how the boundary of the bubble $\yb$ evolves taking into account sources $\lambda$ and $L$.
The coefficients in this equation are
\begin{align}
    &A(\yb) = 1 + S_\ion \yb + S_\ion \Delta + \rho_\ion \yb \Delta,\label{eq:coeffA}\\
    &B(\yb) = \yb \rho_\ion + \frac{S_\ion}{2} + \yb \rho'_\ion \Delta + \rho_\ion \Delta,\\
    &C(\yb) = \frac{1 + (1 + \Delta S_\ion)^2}{2(1+ \Delta S_\ion)^2} S_\ion.\label{eq:coeffC}
\end{align}
Here, $S_\ion \equiv S_\ion(\yb)$, $\rho_\ion \equiv \rho_\ion(\yb)$, $\rho_\ion' \equiv \rho_\ion'(\yb)$.
The coefficients are determined solely by the plasma profile $\rho_\ion(r)$ and the width of the electron sheath $\Delta$.
Interestingly enough, the shape of the electron sheath $g(X)$ does not appear in this equation, unlike in the 3D case.
The sources on the right-hand side are
\begin{align}
    &\lambda(\xi, \yb) = -\int_0^{\yb} {J_x(\xi, y')\dd{y'}},\\
    &L(\xi, \yb) = - \frac{1}{2(1+\Delta S_\ion)} \pdv{}{y} \asquare \bigg|_{y = \yb}.
\end{align}
As the only source of the electric current $J_x$ inside the bubble are the relativistic electron bunches, the first term $\lambda$ describes the influence of the electron driver and accelerated electrons on the shape of the bubble.
Correspondingly, the second term $L$ describes the action of the ponderomotive force of the laser pulse.
Therefore, Eq.~\eqref{eq:bubbleEquationGeneral} allows us to take into account both the driver (either a laser or an electron bunch) and the accelerated electrons when calculating the shape of the bubble.

Typically, a bubble is large compared to the width of the sheath $\yb \gg \Delta$.
However, the width of the sheath is also usually sufficiently large so that $S_\ion \Delta \gg 1$ (see Ref.~\onlinecite{Golovanov_2016_QE_46_295} for additional details for the 3D case).
For example, if uniform plasma is considered, these two conditions correspond to $\yb^{-1} \ll \Delta \ll \yb$.
Under those two conditions, the coefficients \eqref{eq:coeffA}--\eqref{eq:coeffC} are simplified, and Eq.~\eqref{eq:bubbleEquationGeneral} becomes
\begin{equation}
    S_\ion \yb \dv[2]{\yb}{\xi} + \left(\frac{S_\ion}{2} + \yb \rho_\ion \right) \qty(\dv{\yb}{\xi})^2 + \frac{S_\ion}{2} = \lambda + L.
    \label{eq:bubbleEquation}
\end{equation}
The longitudinal electric field can also be found from the shape of the bubble
\begin{equation}
    E_x(\xi) \approx S_\ion(\yb(\xi)) \dv{\yb}{\xi} (\xi)
\end{equation}
We assume that the center of the bubble, i.\,e. the point where it reaches its maximum transverse size, is located at $\xi = 0$, so that the initial conditions are
\begin{equation}
    \yb(0) = y_0, \quad \dv{\yb}{\xi} (0) = 0,
\end{equation}
where $y_0$ is the maximum size of the bubble.
We also assume that there are no sources in the rear part of the bubble ($\xi > 0$), i.\,e. $\lambda = 0$, $L = 0$.
In this case, the solution to Eq.~\eqref{eq:bubbleEquation} for $\xi > 0$ can be found analytically similarly to the 3D case\cite{Golovanov_2016_PoP_23_093114}
\begin{equation}
    \xi = \int_{\yb(\xi)}^{y_0} \frac{\sqrt{y'} S_\ion(y')\dd{y'}}{\sqrt{\int_{y'}^{y_0} S_\ion^2(y'') \dd{y''}  }}.
\end{equation}
This solution defines the function $\yb(\xi)$ implicitly.
It makes it easy to find the half-length of a bubble $\xi_\smax$ by setting $\yb(\xi = \xi_\smax) = 0$.
The electric field in this bubble is 
\begin{equation}
    E_x = - \sqrt{\frac{1}{\yb(\xi)} \int_{\yb(\xi)}^{y_0} S_\ion^2(y') \dd{y'} }
\end{equation}

If the plasma is uniform ($\rho_\ion(y) = 1$, $S_\ion(y) = y$) and there are no sources, Eq.~\eqref{eq:bubbleEquation} becomes
\begin{equation}
    2 \yb \dv[2]{\yb}{\xi} + 3 \qty(\dv{\yb}{\xi})^2 + 1 = 0.
    \label{eq:bubbleEquationUniform}
\end{equation}
It can be compared to the equation for the 3D axisymmetric case (see Ref.~\onlinecite{Lu_2006_PoP_13_056709}) 
\begin{equation}
    \rb \dv[2]{\rb}{\xi} + 2 \qty(\dv{\rb}{\xi})^2 + 1 = 0. \label{eq:bubbleEquation3D}
\end{equation}
While the equation in the 3D case is close to the equation of a circle, Eq.~\eqref{eq:bubbleEquationUniform} resembles the equation of an ellipse $\sqrt{2}$ times longer in the longitudinal direction.
This can be shown by finding a solution to Eq.~\eqref{eq:bubbleEquationUniform} near the center of the bubble ($\xi=0$)
\begin{equation}
    \yb \approx y_0\left(1 - \frac{\xi^2}{4y_0^2}\right)
\end{equation}
which corresponds to an ellipse with semi-axes equal to $\sqrt{2}y_0$ and $y_0$.
However, the electric field in the 2D case
\begin{equation}
    E_x \approx -\frac{\xi}{2}
\end{equation}
is exactly the same as in the 3D case.

\begin{figure}[tb]
    \centering
    \includegraphics[]{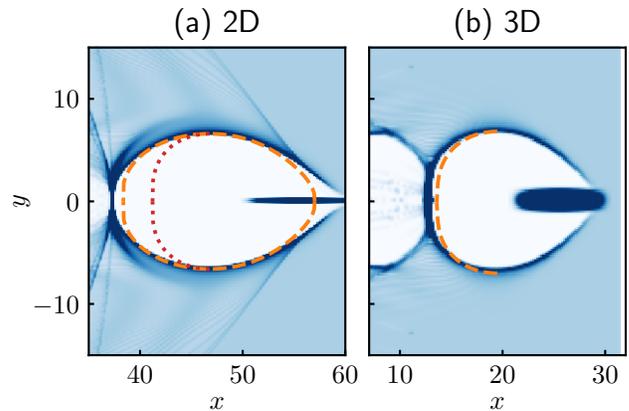}
    \caption{
        Electron density distribution in (a) a 2D bubble, (b) a 3D axysimmetric bubble driven by an electron bunch propagating to the right.
        The dashed lines show the analytic solutions for the boundaries of the bubbles according to Eqs.~\eqref{eq:bubbleEquation} and \eqref{eq:bubbleEquation3D}, respectively.
        The dotted line in (a) shows the analytic solution for a 3D axisymmetric bubble for comparison.
        All lengths are normalized to $c/\omega_\plasm = \lambda_\plasm / 2\pi$.
    }
    \label{fig:bubble2Dvs3D}
\end{figure}

The behavior described above can be observed in particle-in-cell (PIC) simulations.
To demonstrate that, we carried out two-dimensional simulations using the Smilei PIC code. \cite{Smilei, Derouillat_2018_CPC_222_351}
In these simulations, we used an electron bunch driver with the energy of electrons equal to \SI{2}{\GeV}, the maximum charge density of $25n_\plasm$, and the longitudinal and the transverse sizes of $2\lambda_\plasm$ and $0.1\lambda_\plasm$, respectively.
It excited a wakefield in the strongly nonlinear (bubble) regime in uniform plasma.
In Fig.~\ref{fig:bubble2Dvs3D}(a), the resulting electron density distribution in the wakefield and the analytic solution for the bubble's boundary calculated using Eq.~\eqref{eq:bubbleEquation} for the uniform plasma are shown.
For comparison, the analytic solution for the 3D case is drawn with a dotted line.
It is evident that the shape of the bubble in the 2D geometry is closer to an ellipsis stretched in the longitudinal direction than to a circle.
For reference, Fig.~\ref{fig:bubble2Dvs3D}(b) demonstrates a typical spherical bubble of a similar size in the 3D geometry.
The 3D simulations were also performed with the Smilei PIC code.
An electron bunch with the maximum charge density of $40n_\plasm$ and longitudinal and transverse sizes of $1.6 \lambda_\plasm$ and $0.4 \lambda_\plasm$ was used to excite the wakefield in this case.

\begin{figure}[tb]
    \centering
    \includegraphics[]{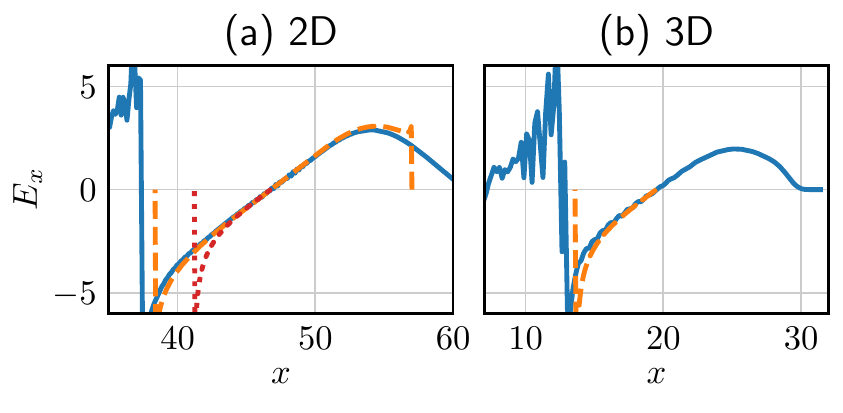}
    \caption{
        Longitudinal electric fields $E_x$ in the bubbles shown in Fig.~\ref{fig:bubble2Dvs3D}.
        The dashed lines correspond to the analytical solutions.
        The dotted line in (a) shows the analytically calculated electric field in the 3D axisymmetric bubble for comparison.
    }
    \label{fig:Ex2Dvs3D}
\end{figure}

The corresponding longitudinal electric fields in the simulations and their comparison to the respective 2D and 3D analytical models are shown in Fig.~\ref{fig:Ex2Dvs3D}.
The simulations support the analytical finding that the dependence of the electric field on the longitudinal coordinate is predominantly linear, and the coefficient of this dependence for uniform plasma is the same for 2D and 3D geometries and is equal to $1/2$.

Figs.~\ref{fig:bubble2Dvs3D} and \ref{fig:Ex2Dvs3D} both show that the developed analytic model fairly accurately describes the bubble observed in the simulations.
The differences occur only at the front and rear edges of the bubble where the assumption that the radial size of the bubble is large becomes incorrect.
Compared to the 3D geometry, a 2D bubble is elongated in the longitudinal direction.
However, the properties of the longitudinal electric field remain the same: it does not depend on the transverse coordinate, is predominantly linear close to the center of the bubble, and its gradient in the uniform plasma is the same as in the 3D case.
This similarity is very important, as the dephasing length, the maximum energy, and the spectra of electrons are determined predominantly by this field.
It might indicate that the resulting properties of the accelerated electron bunches should be qualitatively similar in the 2D simulations compared to the full 3D ones.

\section{Quasi-2D bubble in 3D PIC simulations}
\label{sec:quasi2d}

\begin{figure}[tb]
    \centering
    \includegraphics[]{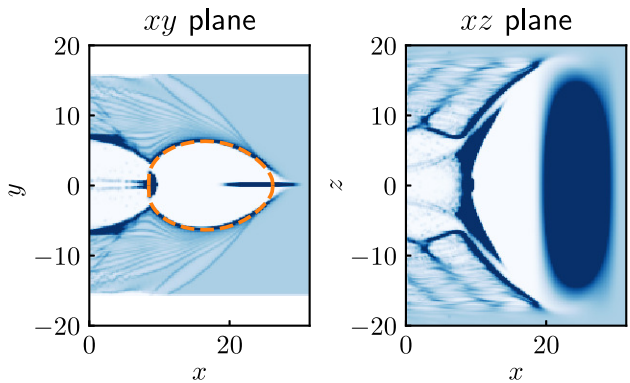}
    \caption{
        Electron density distribution in the $xy$ and $xz$ planes in a bubble excited by a disk-like electron bunch with different transverse sizes.
        The dashed line shows the analytic solution for the two-dimensional bubble in uniform plasma according to Eq.~\eqref{eq:bubbleEquation}.
        All coordinates are normalized to $\lambda_\plasm/2\pi$.
    }
    \label{fig:bubble2d3d}
\end{figure}

In the 3D space, a 2D bubble corresponds to a driver with an infinite size along one transverse direction.
Therefore, it should be possible to create a quasi-2D bubble in the three-dimensional space by using a disk-like driver with one of the transverse sizes significantly exceeding the other.
As an example, a bubble excited by an electron bunch with the maximum charge density of $25 n_\plasm$, the longitudinal size of $\lambda_\plasm$, and the transverse sizes of $0.1\lambda_\plasm$ and $6.4\lambda_\plasm$ along the $y$ and $z$ directions, respectively, is shown in Fig.~\ref{fig:bubble2d3d}.
These parameters correspond to the 2D bubble shown in Fig.~\ref{fig:bubble2Dvs3D}(a).
As the comparison to Fig.~\ref{fig:bubble2Dvs3D}(a) as well as the comparison to the analytical solution (the dashed line in Fig.~\ref{fig:bubble2d3d}) shows, the bubble indeed has the same properties as the 2D bubble in the $xy$ plane.
In the $xz$ plane (corresponding to the plane of the disk-like electron bunch) the bubble has approximately the same size as the driver.

\begin{figure}[tb]
    \centering
    \includegraphics[]{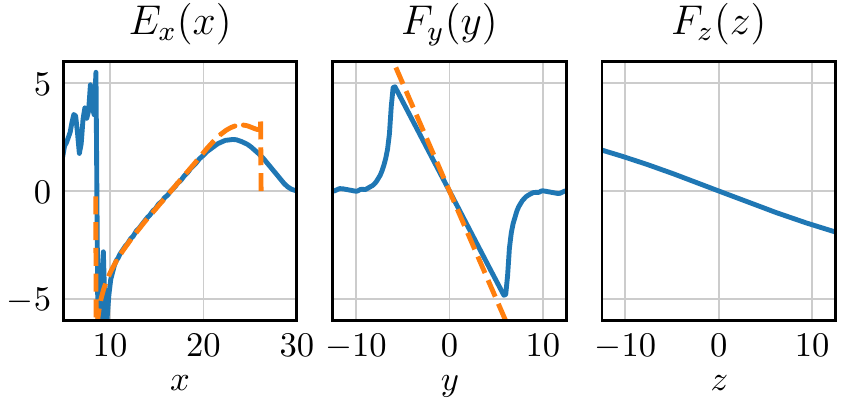}
    \caption{
        The longitudinal electric field $E_x$ on the axis of the bubble and the transverse forces $F_y$ and $F_z$ at $x = 16.5$ in the bubble shown in Fig.~\ref{fig:bubble2d3d}.
        The dashed lines correspond to the analytic solutions.
    }
    \label{fig:fields2d3d}
\end{figure}

The longitudinal electric field and the transverse forces in this bubble are shown in Fig.~\ref{fig:fields2d3d}.
For comparison, the field and the forces predicted by our 2D model are also plotted with the dashed lines.
The comparison shows that the 2D model correctly describes the fields in the bubble.
Obviously, in a 2D bubble of infinite in the $z$ direction size, the transverse force $F_z = 0$.
However, in a quasi-2D bubble, this component is also present.
It is linear in the $z$ direction and focusing for electrons; its gradient is significantly smaller than the gradient of $F_y$.
Therefore, this force should correspond to long-period betatron oscillations in the $z$ direction.

\section{Discussion and conclusions}

We developed a phenomenological model describing the bubble regime of plasma wakefield in the 2D geometry.
In this regime, the influence of the driver (a laser pulse or a relativistic electron bunch) leads to the formation of a cavity free of plasma electrons behind it.
The model is similar to the previous 3D models and is based on the assumption that no plasma electrons are present inside the bubble. 
At the same time, there is a thin electron layer on its boundary.
In the scope of the model, we obtained a differential equation describing the boundary and analytically solved it for absent sources.
The predictions of the model were verified by 2D PIC simulations and showed good correspondence to their result.
In addition, we showed that it is possible to generate a quasi-2D bubble using a disk-like electron bunch in the realistic 3D geometry.
The properties of such a bubble correspond to a bubble observed in 2D PIC simulations.

As 2D simulations are sometimes used as a substitute for more computationally expensive full 3D simulations, 
the most interesting result of the model is the difference in the accelerating and focusing forces in the 2D model compared to a realistic 3D bubble.
The comparison was done both analytically and numerically.
The results show that a bubble in 2D geometry is elongated in the longitudinal direction compared to an  almost spherical bubble in the 3D case.
However, the structure of the forces acting on the electrons inside the bubble remains virtually the same.
The accelerating force is predominantly linear in the longitudinal direction and does not depend on the transverse coordinate; its gradient in uniform plasma is exactly the same as in the 3D case.
The transverse force is also linear and depends only on the transverse coordinate, but its amplitude is two times larger then in the 3D case.
This should significantly affect betatron oscillations and the spectrum of betatron radiation.
Of course, the difference in the wakefield structure is not the only difference introduced by the use of the 2D geometry, as self-focusing of the laser pulse and self-injection and trapping of electrons significantly change as well. \cite {Tsung_2006_PoP_13_056708}
All such differences should be considered when making conclusions from 2D simulations.

\begin{acknowledgements}
The work has been supported by the Russian Science Foundation through Grant No.\,16-12-10383.
\end{acknowledgements}

\section*{References}
\bibliographystyle{aipnum4-1}
\bibliography{Bibliography}

\end{document}